\documentclass[prb,showpacs,aps,superscriptaddress,two column,amsmath,amssymb,floats]{revtex4-1}

\usepackage{hyperref}
\usepackage{graphicx}
\usepackage[usenames,dvipsnames]{color}
\begin{document}

\title{Josephson junction dynamics in the presence of $2\pi$- and $4\pi$-periodic supercurrents}

\author{F. \surname{Dom\'inguez}}
\affiliation{Institut f\"{u}r Theoretische Physik und Astrophysik,Universit\"{a}t W\"urzburg, D-97074 W\"urzburg, Germany}
\author{O. \surname{Kashuba}}
\affiliation{Institut f\"{u}r Theoretische Physik und Astrophysik,Universit\"{a}t W\"urzburg, D-97074 W\"urzburg, Germany}
\author{E. \surname{Bocquillon}}
\affiliation{Physikalisches Institut (EP3), Universit\"{a}t W\"urzburg, D-97074 W\"urzburg, Germany}
\affiliation{Laboratoire Pierre Aigrain, Ecole Normale Sup\'erieure-PSL Research University, CNRS, Universit\'e Pierre et Marie Curie-Sorbonne Universit\'es, Universit\'e Paris Diderot-Sorbonne Paris Cit\'e, 24 rue Lhomond, 75231 Paris Cedex 05, France}
\author{J. \surname{Wiedenmann}} 
\affiliation{Physikalisches Institut (EP3), Universit\"{a}t W\"urzburg, D-97074 W\"urzburg, Germany}
\author{R. S. \surname{Deacon}}
\affiliation{Advanced Device Laboratory, RIKEN, 2-1 Hirosawa, Wako-shi, Saitama, 351-0198, Japan}
\affiliation{Center for Emergent Matter Science, RIKEN, 2-1 Hirosawa, Wako-shi, Saitama, 351-0198, Japan}
\author{T. M. \surname{Klapwijk}}
\affiliation{Kavli Institute of Nanoscience, Faculty of Applied Sciences, Delft University of Technology, Lorentzweg 1, 2628 CJ Delft, The Netherlands}
%\author{H. \surname{Buhmann}}
%\affiliation{Physikalisches Institut (EP3), Universit\"{a}t W\"urzburg, D-97074 W\"urzburg, Germany}
\author{G. \surname{Platero}} 
\affiliation{Instituto de Ciencia de Materiales, CSIC, Cantoblanco, E-28049 Madrid, Spain}
\author{L. W. \surname{Molenkamp}}
\affiliation{Physikalisches Institut (EP3), Universit\"{a}t W\"urzburg, D-97074 W\"urzburg, Germany}
\author{B. \surname{Trauzettel}}
\affiliation{Institut f\"{u}r Theoretische Physik und Astrophysik,Universit\"{a}t W\"urzburg, D-97074 W\"urzburg, Germany}
\author{E. M. {Hankiewicz}}
\affiliation{Institut f\"{u}r Theoretische Physik und Astrophysik,Universit\"{a}t W\"urzburg, D-97074 W\"urzburg, Germany}

\date{\today}

%\pacs{%42.50.Dv,	%Quantum state engineering and measurements
%   73.23.-b,	%Electronic transport in mesoscopic systems
   %03.67.Lx,	%Quantum computation
   %05.60.-k Transport processes
%   05.60.Gg	%Quantum transport
   %74.50.+r, % Tunneling phenomena; point contacts, weak links, Josephson effects
%}

\begin{abstract}
We investigate theoretically the dynamics of a Josephson junction in the framework of the RSJ model. We consider 
a junction that hosts two supercurrrent contributions: a 
$2\pi$- and a $4\pi$-periodic in phase, with intensities $I_{2\pi}$ and $I_{4\pi}$ respectively.
We  study the size of the Shapiro steps as a function 
of the ratio of the intensity of the mentioned contributions, i.e.~$I_{4\pi}/I_{2\pi}$.
We provide detailed explanations where to expect 
clear signatures of the presence of the $4\pi$-periodic contribution
as a function of the external parameters: the intensity AC-bias
$I_\text{ac}$ and frequency $\omega_\text{ac}$.
On the one hand, in the low AC-intensity regime (where $I_\text{ac}$ is much smaller than the critical current, $I_\text{c}$), 
we find that the non-linear dynamics of the junction allows the observation of only even Shapiro steps even in the
unfavorable situation where $I_{4\pi}/I_{2\pi}\ll 1$.
On the other hand, in the opposite limit ($I_\text{ac}\gg I_\text{c}$), even and odd Shapiro steps are present. 
Nevertheless, even in this regime, we find signatures of the $4\pi$-supercurrent in the beating pattern of the even step sizes as a function of $I_\text{ac}$.

\end{abstract}
\maketitle
\section{Introduction}

A topological superconductor forms a new state of quantum matter and possesses a pairing gap in the bulk 
and gapless surface states which in some cases form non-trivial 
Majorana bound states.\cite{Read2000a, Ivanov2001a, Qi2011a}
The Majorana bound states can be interpreted as fermionic particles equivalent 
to their own antiparticles, and have potential applications in fault-tolerant 
topological quantum computation.\cite{Nayak2008a, Alicea2011a, Alicea2012a, Stern2013a}
Additionally to p-wave superconductors like $\rm{Sr}_2\rm{RuO}_4$ or d+id superconductors on hexagonal 
lattices,\cite{Mackenzie2003a, Elster2015a} 
new platforms to host Majorana bound states based on proximitizing ordinary 
singlet-spin superconductor to a material with a strong spin-orbit interaction were proposed.\cite{Kitaev2001a, Fu2009a, Lutchyn2010a, Oreg2010a}
In addition to spectroscopic signatures of the Majorana bound states,\cite{Mourik2012a, Das2012a, Albrecht2016} 
recent experiments on Josephson junctions (JJs) based on Rashba wires or 
topological insulators, which could show topologically non-trivial modes, have attracted a lot of attention.\cite{Rokhinson2012a, Wiedenmann2016a, Bocquillon2016a, Deacon2016a}

%-----------------------------------------------------
Josephson junctions containing a topologically protected Andreev level
exhibit $4\pi$-periodicity in respect to the
superconducting phase difference 
$\varphi$.\cite{Kitaev2001a, Kwon2004a, Fu2009a, Lutchyn2010a, Oreg2010a, Beenakker2013b, Crepin2014a}
Hence, the measurement of topological properties of the JJ involves a probing of the periodicity of the electronic properties
of the junction.
This can be achieved by means of the AC-Josephson effect.\cite{Kitaev2001a}
For example, when the JJ is biased by DC- and AC-currents $I_0+I_\text{ac}\sin(\omega_\text{ac} t)$, 
the average voltage develops plateaus
at integer multiples of 
$\hbar\omega_\text{ac}/2e$, i.e.~$V= n \hbar \omega_\text{ac}/2e$, $n$ being an integer number.\cite{Shapiro1963a} 
These plateaus are known as Shapiro steps and are the result of a 
synchronization process between 
the external driving frequency $\omega_\text{ac}$ and the frequency of the junction $\omega_0$.
Their experimental measurement allows to
establish a direct correspondence between the periodicity of the electronic properties of the
junction and an observable, because when the supercurrent is $4\pi$-periodic  
only even multiples of $\hbar\omega_\text{ac}/2e$ (even Shapiro steps) appear.
The accuracy and universality of this relation has made the Shapiro-steps 
the basis of the international voltage-standard with an accuracy of one part per billion.
Alternatively, one can measure the voltage emission spectrum.\cite{Deacon2016a}
In this case, the $4\pi$-periodicity manifests itself as a resonance line 
separated by the fractional frequency $\omega_0/2$ of the junction.
Nevertheless, these proposals need 
to be performed carefully, due to several side
effects.
For example, relaxation processes may break parity conservation yielding a 
2$\pi$-periodic supercurrent.\cite{Kitaev2001a, Budich2012a, Rainis2012a} 
Furthermore, finite size effects, and the coexistence of the $4\pi$-periodic Andreev state 
together with ordinary Andreev levels with a $2\pi$ periodicity, could 
obscure completely the measurement of the 4$\pi$-periodic 
signal. Proposals based on dynamical transitions allow to overcome 
these difficulties.\cite{Heck2011a, Badiane2011a, San-Jose2012a, Pikulin2012a, Dominguez2012a, Virtanen2013a, San-Jose2013a}
Further proposals circumvent some of these problems by studying
the skewness of the 
$4\pi$-periodic supercurrent profile,\cite{Tkachov2013,Sochnikov2015} or the 
phase-dependent thermal conductance with minimum at $\varphi=\pi$ independent of the 
barrier strength in the heat transport setup.\cite{Sothmann2016} 

During the last years some experiments were performed in JJs where 
the presence of the $4\pi$-periodic Andreev level may be responsible for 
the observations.
In Refs.~\onlinecite{Rokhinson2012a, Wiedenmann2016a, Bocquillon2016a}, 
even Shapiro steps stand alone
at low values of the applied external AC-frequency $\omega_\text{ac}$ and AC-intensity $I_\text{ac}$.
Then, increasing $\omega_\text{ac}$ and/or $I_\text{ac}$, odd Shapiro steps also appear.
A similar phenomenon was observed in Ref.~\onlinecite{Deacon2016a}, where the 
voltage emission spectrum was measured as a function of an external DC-current bias $I_0$. 
For low $I_0$, a signal with the fractional frequency $\omega_0/2$ appears, while 
for increasing $I_0$, one observes a clear transition towards the integer frequency $\omega_0$.
The overall response will be studied on a phenomenological level by the
resistively 
shunted junction (RSJ) model.\cite{Dominguez2012a, Deacon2016a}
Here, we will analyze in detail the dynamics of the RSJ model that carries 
two superconducting contributions $I_{2\pi} \sin(\varphi)$ and $I_{4\pi}\sin(\varphi/2)$, from now on we will call it 2 supercurrents RSJ (2S-RSJ) model.
We will explain the regime of parameters that 
gives rise to the $4\pi\rightarrow 2\pi$ transition and 
explain further signatures arise in the Shapiro step experiment. 

\begin{figure}[tb]
\begin{center}
\includegraphics[width=3.3in,clip]{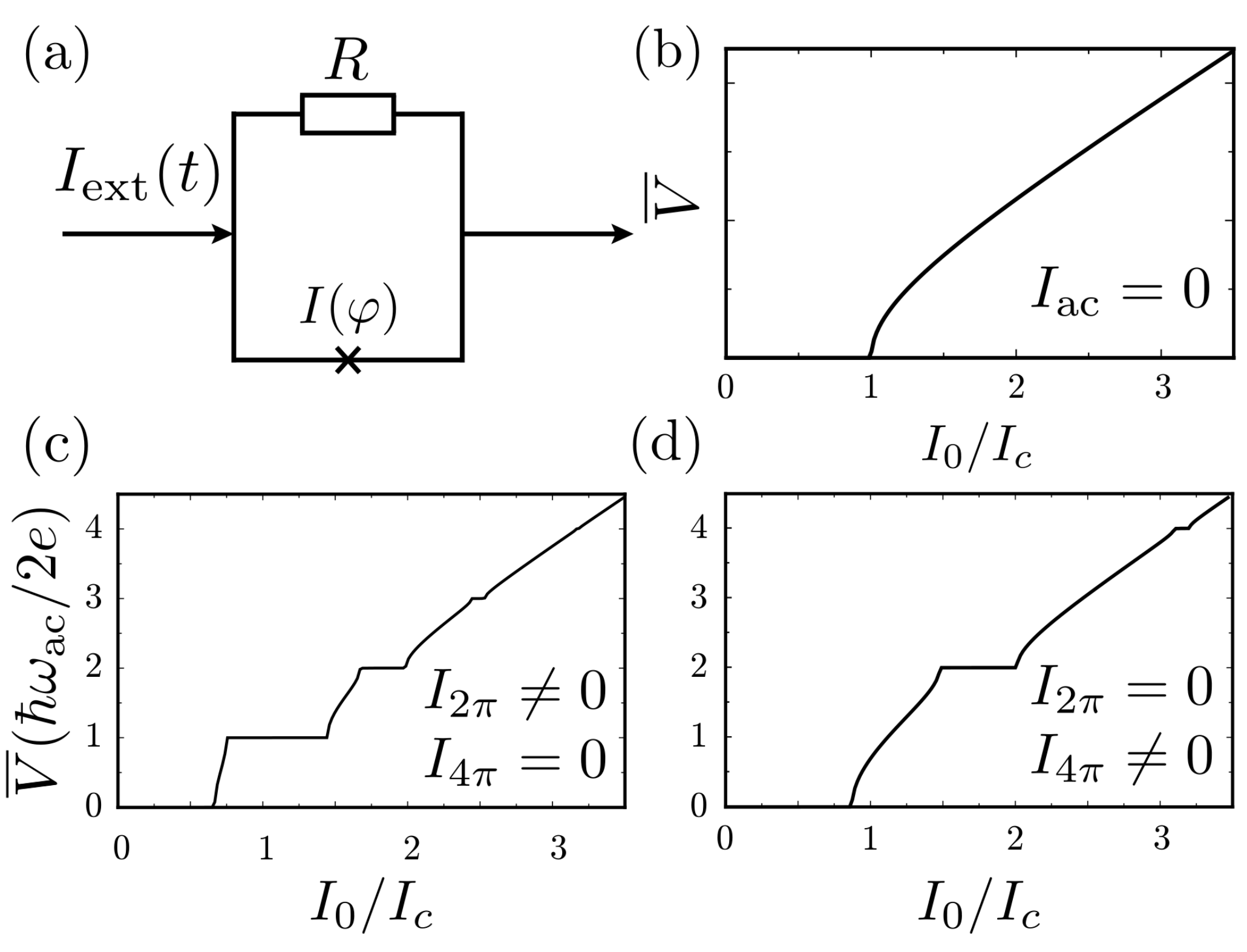}
\end{center}
\caption {\label{Fig.sol4pi}\small (a) Scheme of the RSJ circuit. 
(b) $\overline{V}$ as a function of $I_0$, with $I_\text{ac}=0$. 
The voltage becomes finite for $I_0\geq I_\text{c}$. 
The dependency is $\overline{V}\sim R \sqrt{I_0^2-I_\text{c}^2}$. 
(c, d) We represent the voltage as a function of $I_0$, with $I_\text{ac}\neq0$, and $I_{4\pi}=0$ (c), and $I_{2\pi}=0$ (d).
Thus, the periodicity of the supercurrent is reflected in the parity of the Shapiro steps. 
In panel~c (d), we show Shapiro steps at integer (even) multiples of $\hbar \omega_\text{ac}/2e$.}
\end{figure}

The outline of the paper is as follows. 
In Section \ref{sec:rsj}, we present the
2S-RSJ model---with $2\pi$ and $4\pi$ 
periodic dependence on the phase.
Then, in Sec.~\ref{sec:wp}, we provide a qualitative explanation of the 2S-RSJ model dynamics
by introducing the modified washboard potential (WP). 
In particular, the time-dependent WP allows for a very intuitive 
understanding of the Shapiro step formation as well as reasons for the discrimination 
between the odd and the even steps.
We summarize our knowledge on the non-stationary topological 
Josephson effect in form of a ``phase diagram''.
Finally, in Sec.~\ref{sec:asympt}, we consider
two limits of the 2S-RSJ model, 
the low $I_\text{ac}\ll I_\text{c}$ and the high $I_\text{ac}\gg I_\text{c}$ intensity limits, where $I_\text{c}$ is the critical current of the JJ. 
We solve the 2S-RSJ model analytically in these limits of interest.
In the low intensity limit ($I_\text{ac}\ll I_\text{c}$), we establish the relation
between the emission spectrum experiment and the Shapiro experiment in terms of the
DC-voltage. In addition, we study the step width as a function of $\omega_\text{ac}$.
In the high intensity limit ($I_\text{ac}\gg I_\text{c}$), we explain the beating pattern 
appearing in the even Shapiro step widths as a function of $I_\text{ac}$.

\section{The 2S-RSJ model}
\label{sec:rsj}

\begin{figure}[tb]
\begin{center}
\includegraphics[width=3.31in,clip]{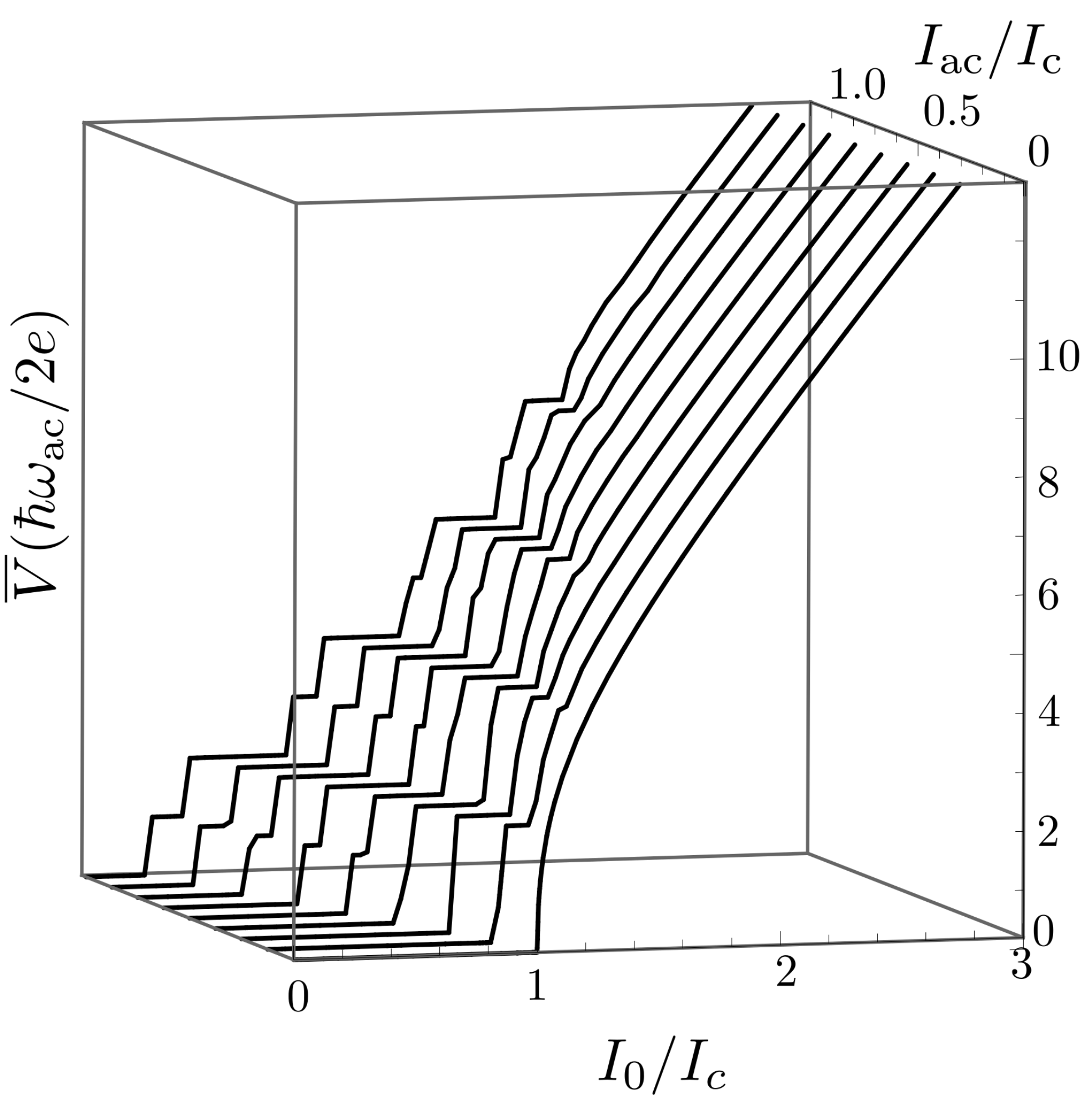}
\end{center}
\caption {\label{Fig.shaphigh}\small 
$I-\overline{V}$ curves for different values of 
$I_\text{ac}=0$ up to $I_\text{c}$, 
with $I_{4\pi}/I_{2\pi}=0.5$, $\omega_\text{ac}=0.2(2eRI_\text{c}/\hbar)$. 
We observe the appearance of odd steps, when $I_\text{ac}\gtrsim I_{4\pi}$.
}
\end{figure}

The RSJ model was introduced 
in Refs.~\onlinecite{Ivanchenko1966b,Stewart1968a, McCumber1968a}.
Under this approach, the JJ dynamics is reduced to the study of an equation of motion,
which can be interpreted as a parallel circuit, including 
three arms: the Josephson junction, a resistive and a capacitive arm.
Here, we will restrict ourselves to the study of the overdamped limit of the 2S-RSJ model, neglecting the capacitive arm, see Fig.~\ref{Fig.sol4pi}(a).\cite{Dominguez2012a,Virtanen2013a} 
This simple model contains the basic ingredients to describe the phase dynamics phenomenologically.
The equation of motion describing the circuit is given by
\begin{align}
&I_\text{ext}(t)= \frac{\hbar}{2eR}\frac{d\varphi}{dt}+I(\varphi),
\label{eq.mainde}
\end{align}
with $I(\varphi)= I_{2\pi} \sin(\varphi)+I_{4\pi} \sin(\varphi/2)$, and 
$I_\text{ext}(t)=I_0+I_\text{ac} \sin(\omega_\text{ac} t)$.
As we explained above, the 4$\pi$-periodic term $I_{4\pi}\sin(\varphi/2)$ 
is of special interest because 
it may originate from the presence of topological superconductivity. 
Writing Eq.~\eqref{eq.mainde} we made several assumptions:
the supercurrent coefficients $I_{2\pi}$ and $I_{4\pi}$ and the resistance $R$ are constant, 
independently of the applied bias $I_\text{ext}(t)$. 
The 2S-RSJ model neglects further dynamical processes such as
quasiparticle poisoning,\cite{Budich2012a,Rainis2012a} or dynamical transitions that might change the
phase periodicity.\cite{Heck2011a,San-Jose2012a,Dominguez2012a, Virtanen2013a}
Furthermore, we expressed the functionality of the supercurrent simply as a sum of two sinusoidal contributions,
which differs from a microscopic derivation. 

The solution of this differential equation provides the
induced voltage $V(t)=\hbar\dot{\varphi}(t)/2e$, 
where $\dot{\varphi}(t)$ is a periodic function with 
period $T_{4\pi}$, and frequency $\omega_0=4\pi/T_{4\pi}$. 
Furthermore,
the average voltage and the frequency are proportional to each other 
by means of $\omega_0=2e\overline{V}/\hbar$, where the overline 
denotes the average over time.

The general features of the current-voltage dispersion 
can be summarized as follows:
Starting from the DC-bias, i.e.~$I_\text{ac}=0$, 
we observe that 
in order to generate a voltage, 
the current bias $I_0$ must exceed the critical value 
$I_\text{c}\equiv$max$\{I(\varphi)\}$ 
[see Fig.~\ref{Fig.sol4pi}(b)].
In this situation, part of the driving current goes through 
the dissipative arm of the circuit and 
therefore a voltage is generated. 
The average voltage can be obtained analytically either for 
$I_{2\pi}=0$ or $I_{4\pi}=0$, and is given by $\overline{V}=R\sqrt{I_0^2-I_\text{c}^2}$.
In the presence of an AC-current the voltage develops Shapiro steps at 
integers multiples of $\hbar \omega_\text{ac}/2e$.
In Figs.~\ref{Fig.sol4pi}(c), and~(d) we show an example of the 
Shapiro experiment only considering $I_{2\pi}$ and $I_{4\pi}$, respectively.
We can see that in the case of a pure $4\pi$- ($2\pi$-) periodic supercurrent, the voltage
contains only even (all) multiples of $\hbar \omega_\text{ac}/2e$.

When both contributions, $I_{2\pi}$ and $I_{4\pi}$, are present
the non-linear dynamics of the
junction governs the low bias regime and gives rise to a very interesting situation:
It is possible to find only even Shapiro steps for a finite range of $I_{\rm ac}$
and even for $I_{4\pi}\ll I_{2\pi}$.\cite{Dominguez2012a}
This phenomenon has been observed experimentally,\cite{Rokhinson2012a, Wiedenmann2016a, Bocquillon2016a} 
and as we will explain below, we can relate it to the
power spectrum of the voltage.\cite{Deacon2016a}
As an example of this,
we show in Fig.~\ref{Fig.shaphigh} 
\emph{I-$\overline{V}$} curves for $I_\text{ac}= 0$ up to $I_{\rm ac}=I_\text{c}$, and $I_{4\pi}/I_{2\pi}=0.5$.
For low values of $I_\text{ac}$ 
we find only even steps, while increasing
$I_\text{ac} \gtrsim I_{4\pi}$, the odd steps emerge.
In the following sections we will present a detailed qualitative and 
quantitative explanations about 
the parameter regime where to expect only even 
Shapiro steps.

\section{The washboard potential}
\label{sec:wp}

We can picture the phase dynamics of the 2S-RSJ model 
as a massless particle sliding on top of a potential,
adapting its velocity instantaneously to its slope.
In order to see this, we rewrite 
Eq.~\eqref{eq.mainde} as $(\hbar/2eR)\dot{\varphi}=-\partial U(\varphi,t)/\partial \varphi$, 
where  
\begin{align}
U(\varphi,t)
=- I_{ext}(t)\varphi+\int d\varphi ~I(\varphi),
\end{align}
is the, so-called, washboard potential.
Here, the external drive term $I_\text{ext}(t)\varphi$
controls the slope, and on top of that, 
the supercurrent contribution modulates the WP profile sinusoidally (see Fig.~\ref{Fig.washboard}a). 
We study the static and dynamical WP, where $I_\text{ac}=0$, and $I_\text{ac}\neq 0$, respectively.

\subsection{Static WP}

\begin{figure}[tb]
\begin{center}
\includegraphics[width=3.4in,clip]{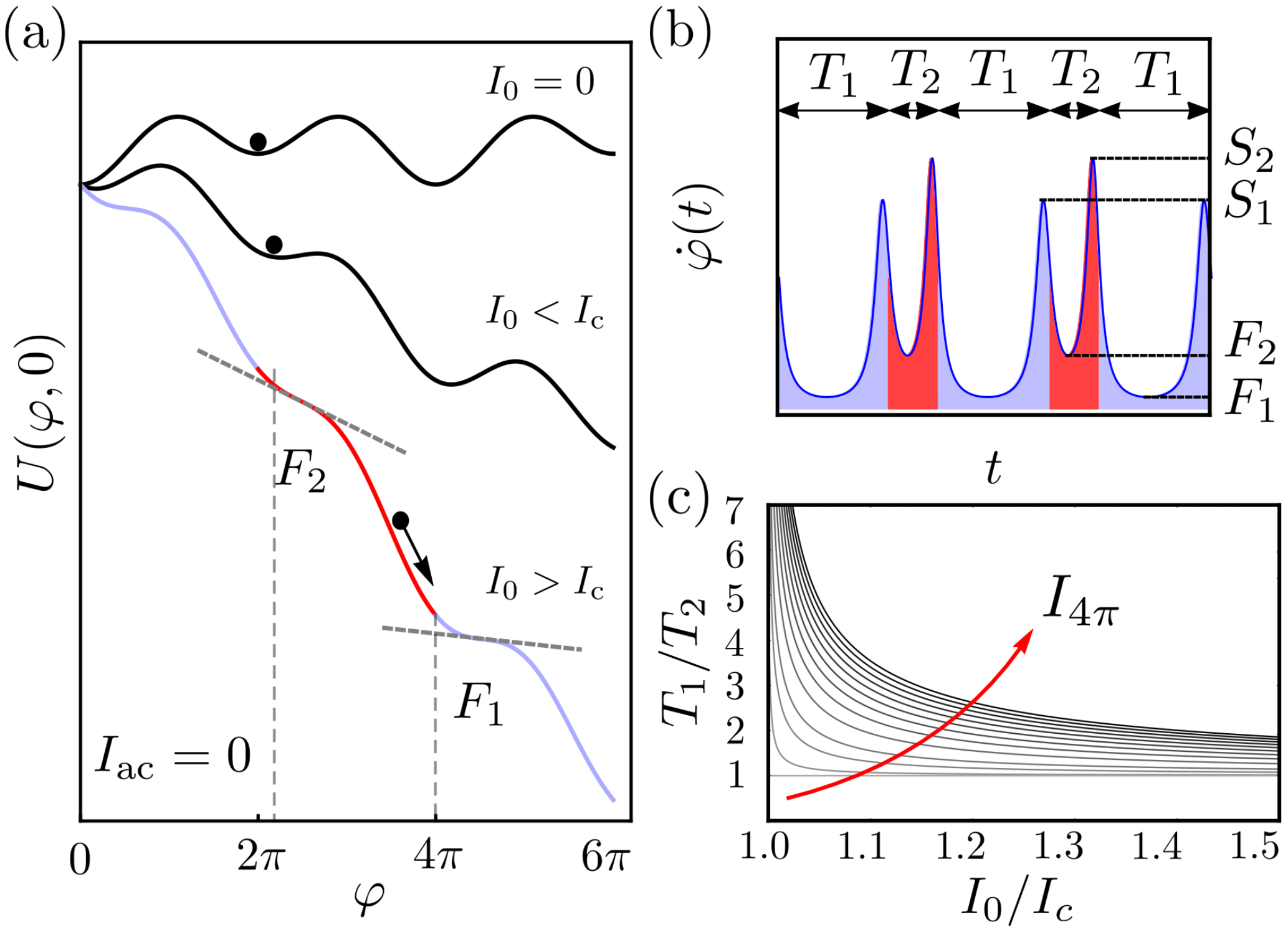}
\end{center}
\caption {\label{Fig.washboard}\small (a) The washboard potential with $I_\text{ac}=0$ as a function of $\varphi$ 
for three different values of $I_0$: top curve $I_0=0$, $I_0<I_\text{c}$ middle curve, and $I_0>I_\text{c}$ bottom curve. 
The dashed lines remark the slope of the WP at the odd ($F_1$) and even ($F_2$) flattest regions. 
We highlight the even and odd sectors in red and blue, respectively.
(b) We show the time evolution of $\dot{\varphi}(\tau)$.\cite{Dominguez2012a} 
We mark in blue (red) the odd (even) sectors according to Eqs.~\eqref{t1} and~\eqref{t2}.
Besides, we can extract from the WP the maxima of $\dot{\varphi}(t)$:
$S_1\approx I_0+I_\text{c}-\sqrt{2}I_{4\pi},~S_2=I_0+I_\text{c},~F_1=I_0-I_\text{c}$ and $F_2\approx I_0-I_\text{c}+\sqrt{2}I_{4\pi}$, being the steepest and the flattest
slopes in each sector, and the equation $(\hbar/2eR)\dot{\varphi}=-\partial U(\varphi,t)/\partial \varphi$ relates the slope and the 
velocity at each time. 
(c) We represent the ratio $T_1/T_2$ as a function of $I_0$ for different values
of $I_{4\pi}/I_{2\pi}$ from zero to one. 
}
\end{figure}

In the absence of AC-bias, the \emph{I-V}-curves exhibit a zero voltage drop
for $I_0\leq I_\text{c}$. This fact is reflected in the WP as minima where the particle rests, 
see Fig.~\ref{Fig.washboard}~(a).
Increasing $I_0$ above the critical value $I_\text{c}$, the local minima in the tilted 
potential vanish, and then, the particle slides along the WP
passing intervals of flatter and steeper slopes.
In this situation, the motion of the particle alternates between slow and rapid sectors. 
We can see the WP profile in Fig.~\ref{Fig.washboard}~(a), and the resulting time evolution 
of $\dot{\varphi}(t)$  in Fig.~\ref{Fig.washboard}~(b), 
characterized by narrow peaks and flat regions.

The presence of the 4$\pi$-periodic contribution
modifies the WP introducing a relative phase between the sectors 
$\varphi_\text{odd}=[4(l-1)\pi,4(l-1/2)\pi]$ and $\varphi_\text{even}=[4(l-1/2)\pi,4l\pi]$, 
$l$ being an integer number.
From now on $\varphi_\text{odd}$ and $\varphi_\text{even}$ will be called odd and even sectors, respectively. 
In the odd sectors, the $4\pi$-term contributes 
with opposite phase to $I_0$ yielding a flatter slope on the WP.
On the other hand, the $4\pi$-current adds to the DC current $I_0$ in the even sectors, and therefore
the slope of the flatter regions become more negative, 
whereas in the odd sectors the $4\pi$-term is subtracted from $I_0$.
We can observe the slope difference between both sectors in Fig.~\ref{Fig.washboard}(a),
where the odd (even) sectors are highlighted in blue (red).
The resulting $\dot{\varphi}(t)$ changes accordingly, and shows different 
maxima depending on the sector parity:
the odd sectors show the steepest and flattest slopes $S_1\approx I_0+I_\text{c}-\sqrt{2}I_{4\pi}$ and $F_1=I_0-I_\text{c}$, respectively, 
while the even sectors $S_2=I_0+I_\text{c}$ and $F_2\approx I_0-I_\text{c}+\sqrt{2}I_{4\pi}$ [see Fig.~\ref{Fig.washboard}(b)]. 
Note, that $S_1$ and $F_2$ are approximate for $I_{4\pi}/I_{2\pi}\ll 1$.

The observed changes of slope cause
differences between the time spent in each sector, which is given by
\begin{align}
&T_{1}=\frac{\hbar}{2eR}\int_0^{2\pi}\frac{d\varphi}{I_0-I_{2\pi}\sin(\varphi)-I_{4\pi}\sin(\varphi/2)}\label{t1},\\
&T_{2}=\frac{\hbar}{2eR}\int_{2\pi}^{4\pi}\frac{d\varphi}{I_0-I_{2\pi}\sin(\varphi)-I_{4\pi}\sin(\varphi/2)},
\label{t2}
\end{align}
where $T_1$ ($T_2$) is the time spent by the particle in the odd (even) sector.
Eqs.~\eqref{t1} and~\eqref{t2} differ on 
the integration range, which introduces a relative sign in $\sin(\varphi/2)$.
In the odd (even) sector $\sin(\varphi/2)$ is always positive (negative), contributing
to a decrease (increase) of the denominator. 
Thus, by construction $T_1\geq T_2$. This is in accordance to the observed
differences between $F_1$ and $F_2$.
Therefore, the ratio $T_1/T_2$ indicates the impact of the 
$4\pi$-supercurrent contribution on the phase dynamics.
For $T_1/T_2\gg 1$ $(T_1/T_2\sim 1)$, 
the particle spends most of the time in the odd (both) 
sectors yielding an effective
4$\pi$ (2$\pi$) WP profile. 
In Fig.~\ref{Fig.washboard}~(c) we plot the ratio $T_1/T_2$ 
as a function of $I_0$, for different values of $I_{4\pi}$.
We observe that for $I_0\sim I_\text{c}$, the ratio $T_1/T_2\gg1$. 
Then, increasing $I_0$ causes a rapid decay of the ratio $T_1/T_2$ towards 1.
Remarkably, we can observe a range of $I_0$ where $T_1/T_2\gg1$, even for very small ratios 
$I_{4\pi}/I_{2\pi}\sim 0.05$.
This means that the junction exhibits a 4$\pi$-periodic dynamics for a finite range of $I_0$.
Obviously, the smaller the ratio $I_{4\pi}/I_{2\pi}$ is, the smaller the range of $I_0$ becomes.
This non-additive phenomenon reveals the highly non-linear dynamics of the 2S-RSJ model.

\begin{figure}[tb]
\begin{center}
\includegraphics[width=3.3in,clip]{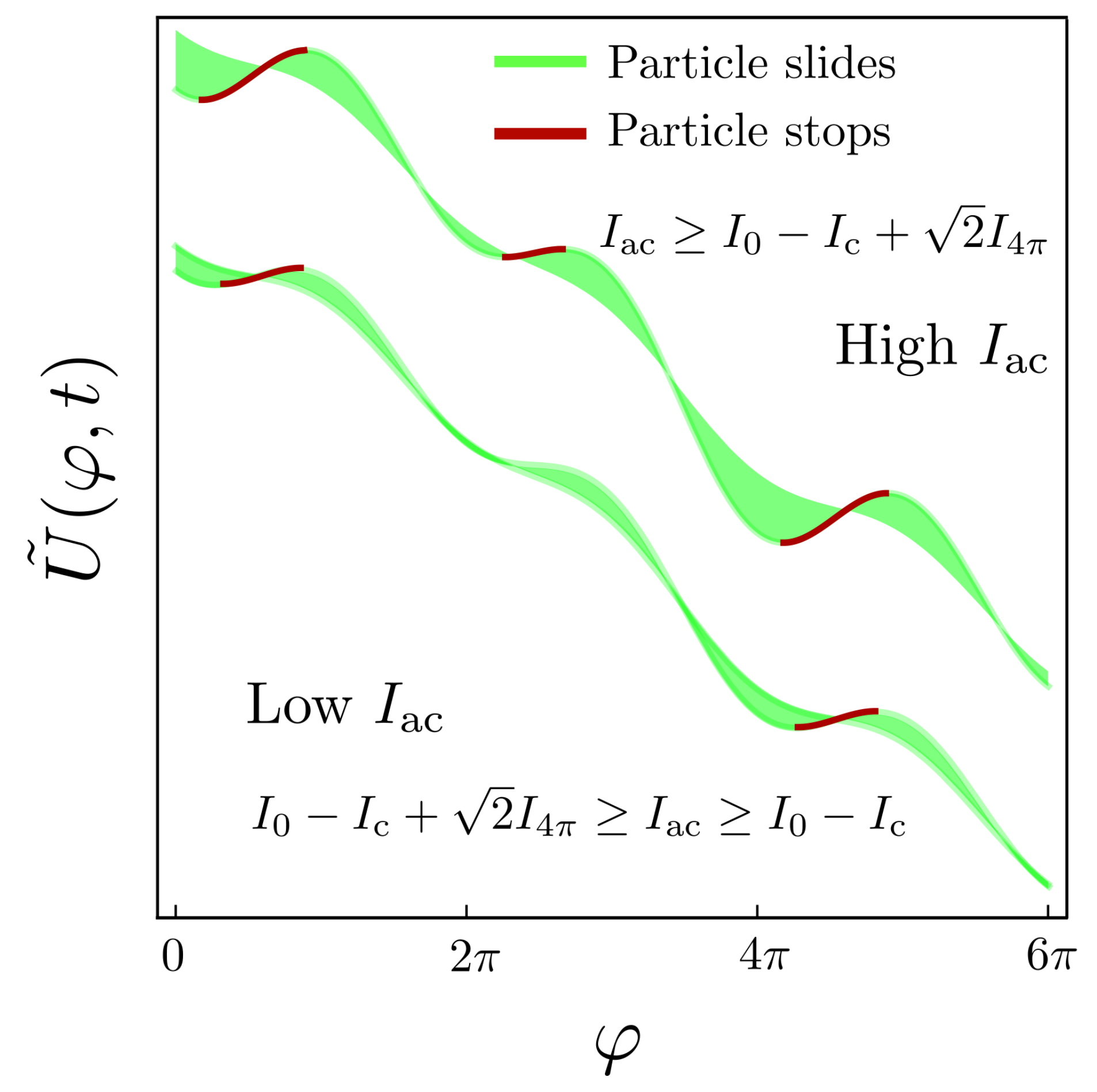}
\end{center}
\caption {\label{Fig.rf}\small 
The renormalized washboard potential $\tilde{U}(\varphi ,t)$ as a 
function of $\varphi$ for two different values of the external bias
$I_0-I_\text{c}<I_\text{ac} < I_0-I_\text{c}+\sqrt{2}I_{4\pi}$ (bottom curve), 
and $I_0-I_\text{c}+\sqrt{2}I_{4\pi}<I_\text{ac}$ (top curve).
We highlight in green the sectors where $\partial \tilde{U} / \partial\varphi<0$ and in red 
$\partial\tilde{U}/\partial\varphi>0$.
}
\end{figure}

We can roughly estimate
$T_1$ and $T_2$  considering that 
the particle spends most of the time in the flattest regions, 
and thus,
$T_1\propto 1/F_1 = 1/(I_0-I_\text{c})$ and 
$T_2\propto 1/F_2\approx 1/(I_0-I_\text{c}+\sqrt{2}I_{4\pi})$.
Note that in the limit of $I_0\gtrsim I_\text{c}$, $T_1$ becomes much larger than
$T_2$. 
In turn $I_0-I_\text{c}\gg I_{4\pi}$, leads to $T_1\sim T_2$.
These considerations on a DC-driven junction explain experimental 
results on the anomalous emission at $\omega_0/2$ of 
topological Josephson junctions,\cite{Deacon2016a} as will be detailed later.

\subsection{Dynamical WP}

The AC-current bias $I_\text{ac} \sin(\omega_\text{ac} t)$ induces a 
time-dependent modulation of the WP slope. 
It enhances 
or reduces the effect of $I_0$ depending on their relative sign. 
At the time periods when $I_0+I_\text{ac} \sin(\omega_\text{ac} t) < I_\text{c}$, 
the current bias recovers the minima, 
where the particle stops.
In order to represent together in a single plot the dynamical WP at different times, 
we show in Fig.~\ref{Fig.rf} a renormalized WP given 
by $\tilde{U}(\varphi, t)=(I_0/|I_\text{ext}(t)|) U(\varphi,t)$, so that $U$ and $\tilde{U}$ coincide for $I_{\rm{ac}}=0$.
Thus, we separate visually the average tilting 
from the AC-bias slope, while we keep the 
local sign of the slope unchanged at any time.
The regions with positive slope (marked red) are 
impenetrable for the particle at the given moment of time.
The periodic appearance of the red intervals realizes a turnstile mechanism,
which allows the phase to propagate an integer multiple $m$ of green intervals between the minima
per cycle.
This manifests itself in the relation $\omega_{0}=n\omega_\text{ac}$, where
the particle slides through $m$ green intervals of total length $2\pi n$ 
until it stops.
Shapiro step arises if the resonance (with fixed $n$ and $m$) holds for a finite range of $I_0$.
This means that the different tilting $I_0$ of the WP is compensated by the stopping periods. Thus, 
the particle's average speed ($\langle \dot{\varphi}\rangle$) remains constant.
%In this situation 
%This mechanism differs from an ordinary
%resonance mechanism, because it takes place over a finite range of $I_0$

\begin{figure}[tb]
 \begin{center}
 \includegraphics[width=\columnwidth]{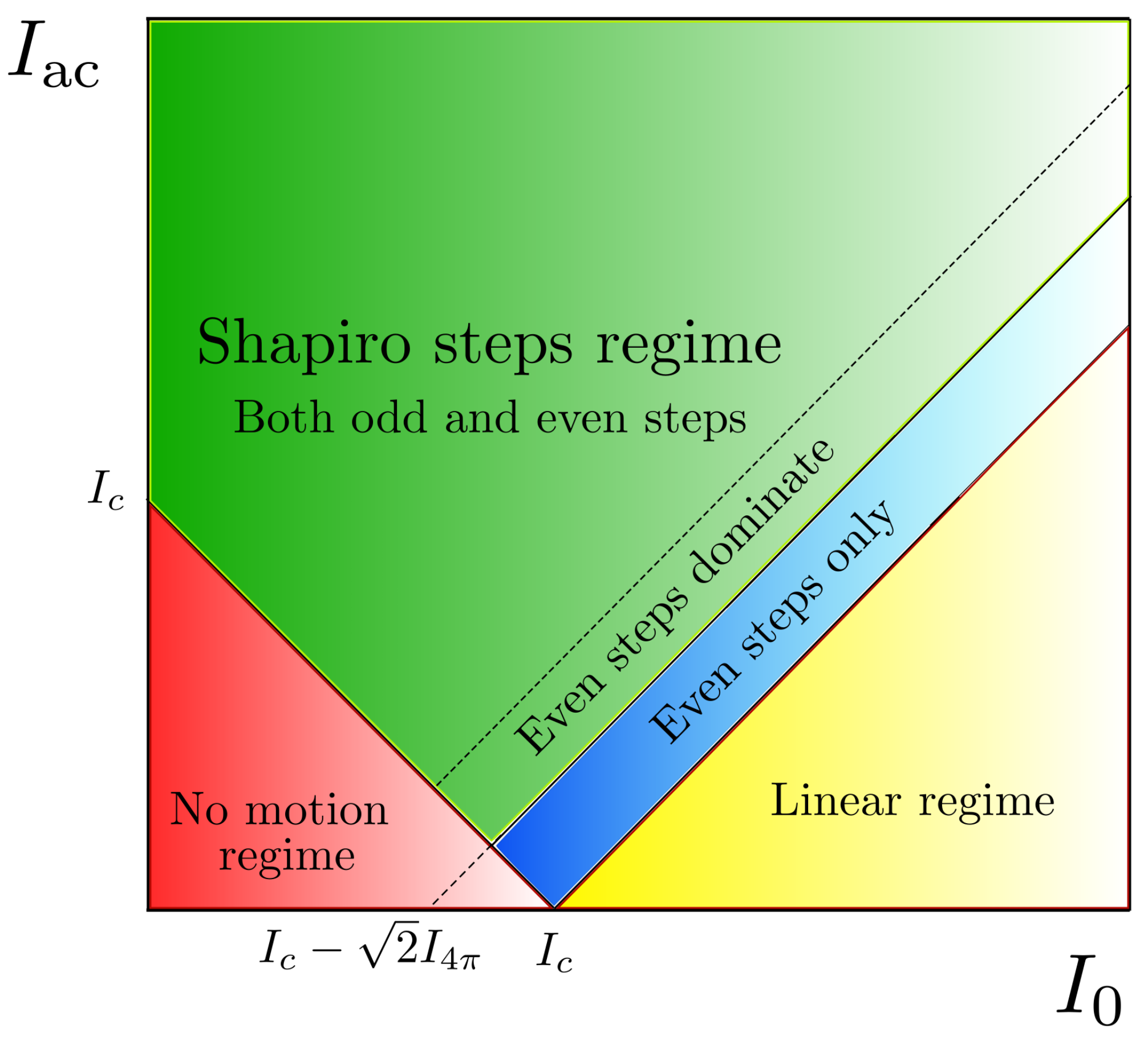}
 \end{center}
 \caption {\label{Fig.diagram}\small 
Phase diagram of the voltage as a function of $I_0$ and $I_\text{ac}$. 
We differentiate between ``No motion regime'' (red area), where $\overline{V}=0$.
The ``linear regime'' (yellow area), where there is no Shapiro steps but $\overline{V}\neq0$.
Finally, the ``Shapiro steps regime'' (green and blue areas).
Remarkably, following the WP considerations we expect to observe only even steps 
in the blue area, i.e.~for $I_0-I_\text{c}+\sqrt{2}I_{4\pi}\gtrsim I_\text{ac}\gtrsim I_0-I_\text{c}$.
}
\end{figure}

Interestingly, 
we can find a situation where only the $4\pi$-contribution becomes visible. 
When the AC-current is set such that it fulfills $|F_2| \gtrsim I_\text{ac} \sin(\omega_\text{ac}t) \gtrsim |F_1|$,
the WP recovers temporarily the minima in the odd sectors only, being separated by a phase difference of $4\pi$ and not $2\pi$ (see bottom curve in Fig.~\ref{Fig.rf}).
Thus, the periodicity of the junction is effectively that of a pure $4\pi$-periodic one. 
Hence, we expect to observe only even Shapiro steps since $2\pi n =4\pi m$. 
On the other hand, for the period of time where 
$ |F_2|\lesssim I_\text{ac} \sin(\omega_\text{ac}t)$, 
the particle is temporarily stopped 
at each sector, yielding any multiple of Shapiro step (see top curve in Fig.~\ref{Fig.rf}).

We summarize these qualitative results in 
Fig.~\ref{Fig.diagram}, 
where we estimate the parameter regime of the Shapiro steps 
as a function of $I_\text{ac}$ and $I_0$. 
We differentiate between three regimes: 
``No motion regime'' (red area), limited by $I_0+I_\text{ac}<I_\text{c}$. Here, the WP exhibits 
always minima where the particle rests, yielding a zero average voltage $\overline{V}=0$.
The ``Linear regime'' (yellow area) extends over $I_0-I_\text{c}>I_\text{ac}$, where the WP cannot stop the particle
at any time, yielding a finite voltage without developing steps. 
Finally, the ``Shapiro steps regime'' (green and blue areas), is the region limited by 
$I_0-I_\text{c}<I_\text{ac}$ and $I_0+I_\text{ac}>I_\text{c}$. 
Following the arguments presented above, we 
distinguish an inner blue region 
$I_0-I_\text{c}+\sqrt{2}I_{4\pi}\gtrsim I_\text{ac}\gtrsim I_0-I_\text{c}$ where 
we expect to observe the even steps only. Increasing further $I_\text{ac}$, we expect to 
observe a crossover where odd steps appear together with even steps,
with a dominating even steps contribution.
Then, for $I_\text{ac} > I_0-I_\text{c}+\sqrt{2}I_{4\pi} $
we expect to have even and odd Shapiro steps, without any clear dominance.

We now understand the underlying reason for
the observation of even Shapiro steps.
However, we have not discussed so far the role 
played by $\omega_\text{ac}$ in this phenomena, which is the subject of the next section.
Indeed, it is known that the effect of increasing the value of $\omega_\text{ac}$ has a similar 
effect as increasing $I_\text{0}$.\cite{Dominguez2012a}
This can be understood in the following way: since the Shapiro steps occur at 
$\omega_0= n\omega_\text{ac}$, where $n \in \mathbb{N}$,  and thus,
tuning $\omega_\text{ac}$ requires the change of $\omega_0$, which is also externally tuned by $I_0$.
Nevertheless, this reasoning is vague and deserves a quantitative study.
Therefore, in order to understand this phenomenon we perform a perturbative approach 
to the equation of motion in the next section.

\section{Asymptotic limits of the 2S-RSJ model}
\label{sec:asympt}

We study two asymptotic limits of the 2S-RSJ model that have experimental relevance. 
First, the low intensity limit $I_\text{ac}\ll I_\text{c}$ is the limit where we can expect to
observe only even Shapiro steps even
for $I_{4\pi}/I_{2\pi}\ll 1$.
Second, the high intensity limit, 
$I_\text{ac}\gg I_\text{c}$, where both steps are present.
Before entering into the study of the asymptotic limits, it is convenient to
rewrite Eq.~\eqref{eq.mainde} using dimensionless units. 
We first
divide Eq.~\eqref{eq.mainde} by the critical current $I_\text{c}$. 
Then, we make the change of variable
\begin{align}
\tilde{t}=(2eRI_\text{c}/\hbar)t,\nonumber
\end{align}
and substitute currents and frequencies as follows
\begin{align}
\tilde{I}_i= \frac{I_i}{I_\text{c}},\text{~~}\tilde{\omega}_\text{ac}=\frac{\hbar\omega_\text{ac}}{2eR I_\text{c}}.\nonumber
\end{align}
Then, Eq.~\eqref{eq.mainde} yields
\begin{align}
\tilde{I}_{0}+\tilde{I}_\text{ac} \sin(\tilde{\omega}_\text{ac} \tilde{t})=\frac{d\varphi}{d \tilde{t}}+\tilde{I}_{2\pi}\sin(\varphi)+\tilde{I}_{4\pi}\sin(\varphi/2).
\label{eomdl}
\end{align}
In this notation the critical current is normalized to $1$,
namely
\begin{align}
\tilde{I}_\text{c}=1=\text{max} \{ \tilde{I}_{2\pi}\sin(\varphi)+ \tilde{I}_{4\pi}\sin(\varphi/2)\}.
\end{align}
Derived quantities such as the voltage or the frequency of the junction are given by $\tilde{\overline{V}}=\overline{V}/I_\text{c} R$ and 
$\tilde{\omega}_{0}=\hbar\omega_{0}/2eR I_\text{c}$, respectively.
Thus, the Josephson relation is $\tilde{V}=\tilde{\omega}_0$, showing that 
the voltage and the frequency of the junction are equal.

In order to keep the notation as simple as possible, from now on we 
skip the tildes, implying the dimensionless variables, and restore dimensionality in the conclusions.
In these new units we will study: the low 
($I_\text{ac}\ll 1$) and the high intensity limits ($I_\text{ac}\gg 1$).

\subsection{Low intensity limit: $I_\text{ac}\ll 1$}

In this limit we treat the 
AC-driving as a perturbation, thus, we expand
$\varphi (t)$ in powers of 
$I_\text{ac}$,\cite{Aslamazov1969a,Thompson1973a} that is
\begin{align}
\varphi= \varphi_0+I_\text{ac} \,\varphi_1+I_\text{ac}^2 \,\varphi_2+\cdots.\nonumber
\end{align}
The zeroth-order contribution $\varphi_0$ corresponds to the DC-driven solution of the 2S-RSJ equation 
and the $\varphi_n$ is the $n^{\textrm th}$-order correction.
In this limit the width of the Shapiro steps is proportional to $I_\text{ac}$.
In order to determine their width we perform a trick\cite{Aslamazov1969a,Thompson1973a} which consists of splitting
$I_0$, which is a constant parameter into
\begin{align}
I_0=I_\text{v}+I_\text{ac}\,\beta_1+I_\text{ac}^2\,\beta_2+\cdots\nonumber
\end{align}
Here, $I_\text{v}$, is given by the value of $I_0$ at the beginning of the step. The rest of the terms ($\beta_n$)
leave constant the voltage.
In this way, the zeroth-order contribution determines the voltage 
$\langle \dot{\varphi}\rangle=\langle \dot{\varphi}_0\rangle$, yielding  
$\langle \dot{\varphi}_n\rangle=0$, for $n\neq0$. 
Therefore, we need to determine $\beta_n$
that cancels the $n^{\textrm th}$-order contribution of the voltage, 
i.e.~$\langle \dot{\varphi}_n\rangle=0$. 
As we will see below, this gives the step width:
the range of $I_0$ in which the voltage remains constant.

\subsubsection{Zeroth-order contribution in $I_{\rm{ac}}$: Power spectrum}

Using the above definitions we obtain the 
zeroth-order differential equation
\begin{align}
I_\text{v}= \dot{\varphi}_0+I_{2\pi}\sin(\varphi_0)+I_{4\pi} \sin(\varphi_0/2).
\label{eq.a1low}
\end{align}
Its exact analytical solution is 
cumbersome and does not provide 
any further insight with respect to 
the numerical solution. 
For this reason, we have 
adapted the solution of a 2$\pi$-junction\cite{Aslamazov1969a} taking into account 
the presence of two periods, $T_1$ and $T_2$ given in Eqs.~\eqref{t1} and~\eqref{t2}, and adjusting the intensity 
of the function.
See further details in App.~\ref{app.2pi}. 
Doing so we obtain
\begin{equation}
\begin{split}
\dot{\varphi}_0(t)\approx & \,\omega_0 \left[1+\right.\\
& \sum_{n=1}^\infty  z^{n} \left(2\cos(n \omega_0 T_1/4)\cos(n \omega_0 t/2)+\right.\\
&\left. \left.(I_{2\pi}-1)\sin(n\omega_0 T_1/4)\sin(n \omega_0 t/2)\right)\right].
\end{split}
\label{phidot}
\end{equation}
Besides, 
the amplitudes of the harmonics decrease 
in geometric progression with $z=\sqrt{I_\text{v}-\omega_0}$. 
This approximation shows the numerical solution coming out 
from Eq.~\eqref{eq.a1low} (see App.~\ref{app.2pi}), specially for $I_{4\pi}/I_{2\pi}\leq 0.5$. 
\begin{figure}[tb]
\begin{center}
\includegraphics[width=3.2in,clip]{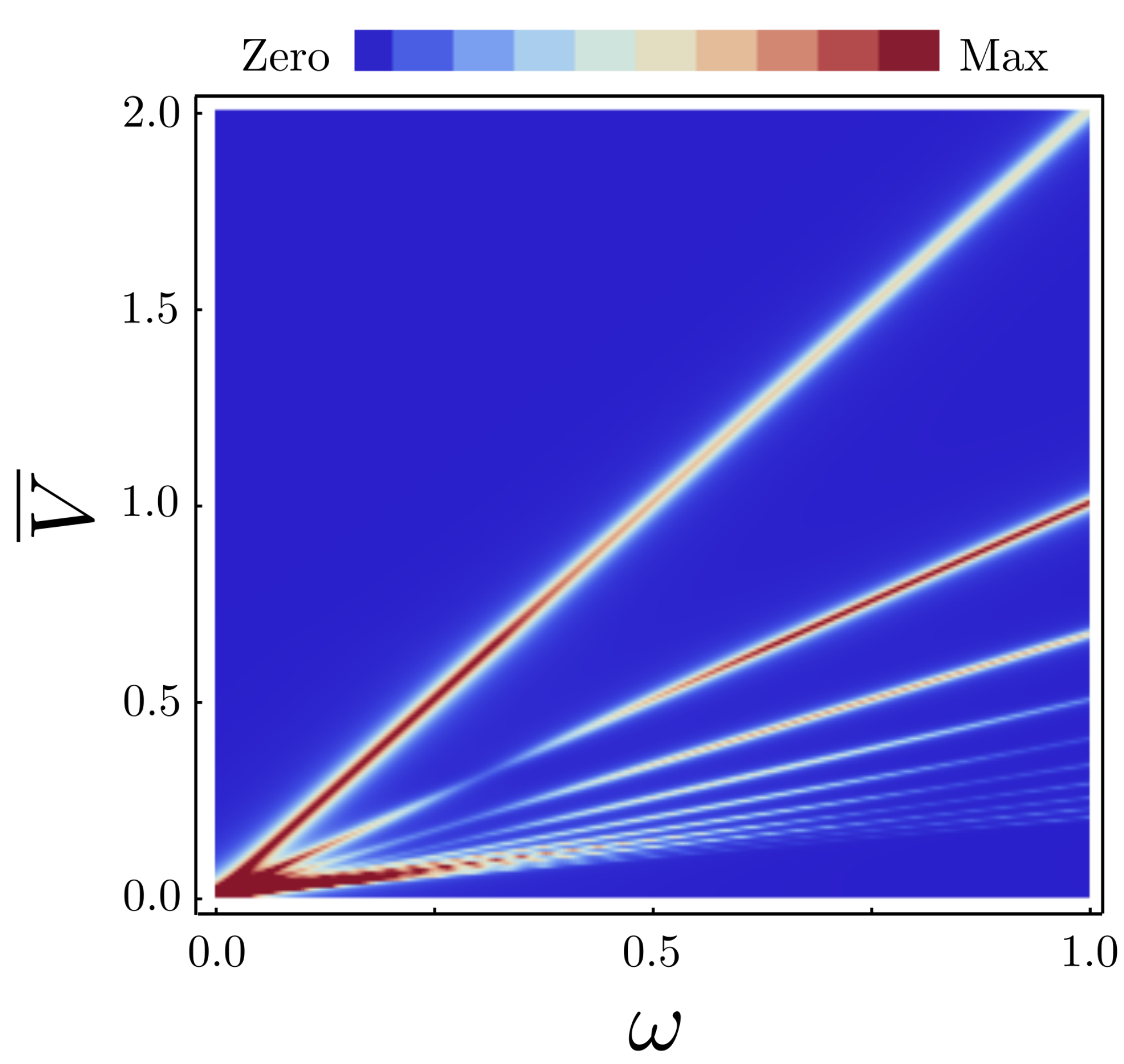}
\end{center}
\caption {\label{Fig.fourier}\small 
Low intensity limit $I_\text{ac}\ll 1$: 
Fourier transform (colorscale) $\dot{\varphi}_0(\omega)=|\int dt e^{i\omega t} \dot{\varphi}_0(t)|$ 
as a function of $\omega$ and the voltage $V=\omega_0$. 
The intensity of the resonances follow Eq.~\eqref{phi0}.
The first two resonance lines with higher slope correspond 
to the frequencies $\omega=\omega_0/2$ (fractional frequency) and $\omega=\omega_0$.
The rest of the resonance lines correspond to higher harmonics.
}
\end{figure}
The Fourier transform of Eq.~\eqref{phidot} is proportional to 
the emission spectrum of the voltage, and has been measured 
in Ref.~\onlinecite{Deacon2016a}.
Performing the Fourier transform of Eq.~\eqref{phidot}, 
$\dot{\varphi}_0(\omega)=|\int dt e^{i\omega t} \dot{\varphi}_0(t)|$ we obtain
\begin{equation}
\begin{split}
\dot{\varphi}_0(\omega)\approx&\delta(\omega-n\omega_0/2) z^{n}\omega_0 \left[4\cos^2\left(\frac{n T_1}{T_1+T_2} \pi\right)+\right. \\
&\left. (I_{2\pi}-1)^2\sin^2\left(\frac{n T_1}{T_1+T_2} \pi\right)\right]^{1/2},
\label{phi0}
\end{split}
\end{equation}
where the delta function $\delta(\omega-n\omega_0/2)$ makes $\dot{\varphi}_0(\omega)$
finite for $\omega=n\omega_0/2$, with $n=1$ ($n=2$) giving the
fractional (integer) frequency $\omega_{0}/2$ ($\omega_{0}$).
Here,
we have made use of the relation $\omega_0=4\pi/(T_1+T_2)$. 

In Fig.~\ref{Fig.fourier} we represent $\dot{\varphi}_0(\omega)$ as a function of $\omega$ and $V=\omega_0$.
We will focus on the two top resonance lines, which correspond from top to bottom to the
frequencies $\omega_0/2$ ($n=1$, i.e.~$\omega=\omega_0/2$) and $\omega_0$ ($n=2$, i.e.~$\omega=\omega_0$), respectively. 
We can observe that the fractional contribution
with $n=1$ [$\dot{\varphi}_0(\omega_0/2)$] 
dominates over the $2\pi$-contribution with $n=2$ [$\dot{\varphi}_0(\omega_0)$] for low values of $\omega_0$. 
Increasing further $\omega_0$, this tendency is reversed and the $2\pi$-contribution dominates. 
As we explained above, this can be understood in terms of 
the ratio $T_1/T_2$, which decreases as a function 
of $I_0$, as it was shown in Fig.~\ref{Fig.washboard}(c) 
(note that $\omega_0$ is tuned by $I_0$).

For simplicity, we analyze the limit where $I_{2\pi}\gg I_{4\pi}$, which
yields in our dimensionless units $I_{2\pi}\sim 1$ making
the second term in Eq.~\eqref{phi0} negligible.
In this scenario, the coefficient 
$\cos(n \pi T_1/(T_1+T_2))$ rules the periodicity of the voltage. 
In the limit where $T_1\gg T_2$, $\cos^2(n \pi T_1/(T_1+T_2))\approx 1$,
and the Fourier expansion contains only one frequency, i.e.~$\omega_0/2$ 
and its harmonics. 
Therefore, the junction behaves like a pure $4\pi$-periodic junction. 
In the opposite limit where $T_1\sim T_2$, the arguments $T_1/(T_1+T_2)\approx 1/2$, 
thus, Eq.~\eqref{phi0} only contains even terms, and thus, 
the frequency $\omega_0/2$ is doubled to $\omega_0$, yielding a $2\pi$ contribution.
This $4\pi\rightarrow 2\pi$ transition is shown in Fig.~\ref{Fig.fourier} and is consistent with the 
emission spectrum experiment performed in Ref.~\onlinecite{Deacon2016a}. 
The value of $\omega_0$ at which the integer 
contribution $n=2$ overcomes the fractional contribution $n=1$
depends only on the ratio $I_{4\pi}/I_{2\pi}$. Thus, a direct comparison with the experimental 
results provides the value $I_{4\pi}$.\cite{Deacon2016a}

\subsubsection{First-order contribution in $I_{\rm{ac}}$: Shapiro steps width}

The first order contribution is obtained from the solution of
the linear differential equation
\begin{align}
\beta_1+\sin(\omega_\text{ac} t)= \dot{\varphi}_1+\varphi_1\left(I_{2\pi}\cos(\varphi_0)+\frac{I_{4\pi}}{2} \cos(\varphi_0/2)\right),
\label{eq.first}
\end{align}
which can be solved using the integrating 
factor $\exp(\int dt \left(I_{2\pi}\cos(\varphi_0)+I_{4\pi}/2 \cos(\varphi_0/2)\right))$.
At this point it is particularly useful to realize that
\begin{align}
I_{2\pi}\cos(\varphi_0)+\frac{I_{4\pi}}{2} \cos(\varphi_0/2)=-\frac{\ddot{\varphi}_0}{\dot{\varphi}_0}.
\end{align}
\begin{figure}[tb]
\begin{center}
\includegraphics[width=3.2in,clip]{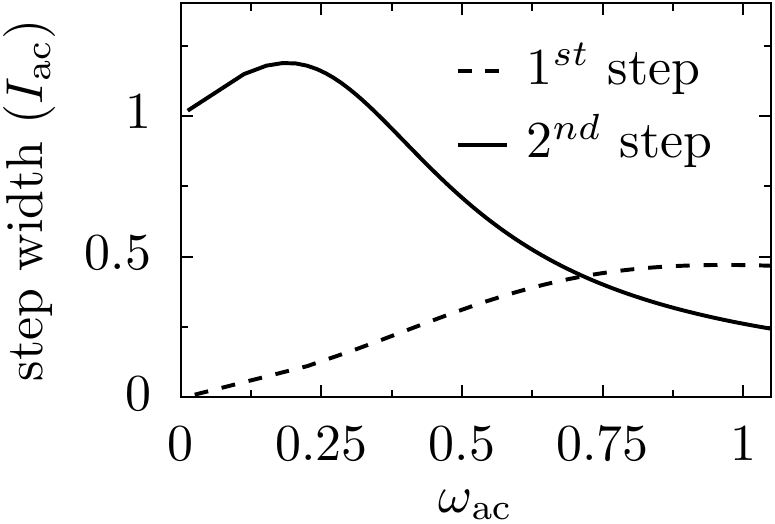}
\end{center}
\caption {\label{Fig.shap}\small 
Low intensity limit $I_\text{ac}\ll 1$: First (dashed curve) and second 
(solid curve) Shapiro steps width in units of $I_\text{ac}$ as a function of $\omega_\text{ac}$.
The width of the Shapiro steps is calculated from Eq.~\eqref{stwidth}.
}
\end{figure}
This relation simplifies greatly Eq.~\eqref{eq.first}, yielding
\begin{align}
\varphi_1(t)=\dot{\varphi}_0(t)\int_0^ t dt' (\beta_1+\sin(\omega_\text{ac} t'))\frac{1}{\dot{\varphi}_0(t')}.
\label{eq.first2}
\end{align}
In order to extract the width of the first two Shapiro steps we 
need to find the value of $\beta_1$ that makes $\langle \dot{\varphi}_1\rangle=0$, that is,
$\varphi_1(T)/T=0$, where $T\rightarrow \infty$. 
This involves the cancellation of 
the constant terms in the integrand of Eq.~\eqref{eq.first2}.
The rest of the terms are canceled by the factor $1/T$.
Thus, when $\omega_\text{ac}=n\omega_0/2$ we find the equality 
\begin{align}
\beta_1  f_0 + f_n \exp\left( i (\omega_\text{ac}-n \omega_0/2) t \right)=0,
\label{eq.condition}
\end{align}
where $f_n$ are the Fourier coefficients of $1/\dot{\varphi}_0(t)$, 
namely,
\begin{align}
\frac{1}{\dot{\varphi}_0(t)}=\sum_{n=-\infty}^\infty f_n \exp\left( i n \omega_0/2 t\right).
\end{align}
The solution for $n=1$ corresponds to 
the second step $(\omega_0=2\omega_\text{ac})$, while for $n=2$ to the first step $(\omega_0=\omega_\text{ac})$.
The step width is given by the equation
\begin{align}
\beta_1(n\omega_0/2)=2\left|\frac{f_n}{f_0}\right|.
\label{stwidth}
\end{align}
Note that in pure 2$\pi$-junctions the first order contribution only contains 
solutions for the first step width. 
In turn, when both contributions are present, the first-order contribution 
provides the width of the first and the second steps.
In Fig.~\ref{Fig.shap} we show the value of $\beta_1(n\omega_0/2)$ (the step width) as a 
function of $\omega_\text{ac}$.
We can observe that the second step 
dominates for low values of $\omega_\text{ac}$, and decreases at higher values.
This behavior is rather similar to the one observed in the power spectrum, 
where for $I_0-I_\text{c}\lesssim \sqrt{2}I_{4\pi}$, the fractional signal is more visible.
Therefore, we can establish the connection 
between the periodicity of the Shapiro experiment and the radiated power 
spectrum observed in Refs.~\onlinecite{Rokhinson2012a, Wiedenmann2016a, Bocquillon2016a, Deacon2016a}, 
since we see from Eq.~\eqref{stwidth} that the Shapiro steps are proportional to the Fourier transform of $1/\dot{\varphi}_0(t)$.

\begin{figure}[tb]
\begin{center}
\includegraphics[width=3.3in,clip]{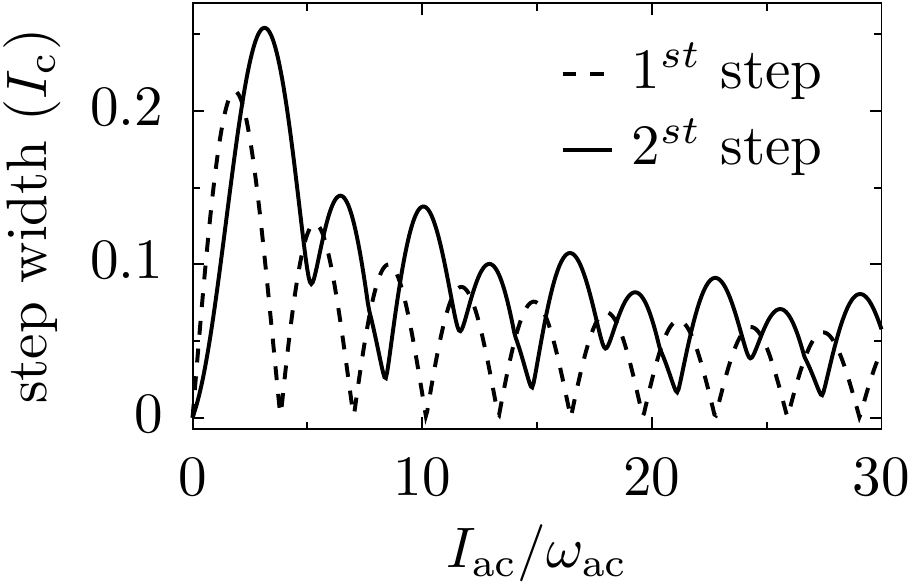}
\end{center}
\caption {\label{Fig.bessel}\small 
High intensity limit $I_\text{ac}\gg 1$: Width of the first two steps as a function of $I_\text{ac}/\omega_\text{ac}$. 
The oscillatory behavior is due to the Bessel functions (see Eqs.~\ref{n1} and \ref{n2}). Interestingly, even Shapiro steps 
exhibit a beating pattern produced by the coexistence of $2\pi$ and $4\pi$ supercurrents.}
\end{figure}

\subsection{High intensity limit: $I_\text{ac}\gg 1$}

In this limit the zeroth-order contribution is obtained neglecting the supercurrent contributions, thus
\begin{align}
I_0+I_\text{ac}\sin(\omega_\text{ac} t)=\frac{d\varphi_0(t)}{dt},
\label{hil0}
\end{align}
where $\varphi_0(t)$ is the zeroth contribution, in units of $I_\text{c}$.
Eq.~\eqref{hil0} can be integrated exactly, 
\begin{align}
\varphi_0(t)=I_0 t -\frac{I_\text{ac}}{\omega_\text{ac}}\cos(\omega_\text{ac} t)+\phi_0,
\label{sol0}
\end{align}
where $\phi_0$ is a constant phase that needs to be determined, see below.
Since we have linearized the differential equation, the average voltage at zeroth order is
$\langle\dot{\varphi}_0\rangle=I_0$.
In order to recover the Shapiro steps
we need to take into account the first order contribution, given by
\begin{align}
\frac{d\varphi_1(t)}{d t}=-I_{2\pi}\sin(\varphi_0(t))-I_{4\pi} \sin(\varphi_0(t)/2).
\label{eq1order}
\end{align}
$\dot{\varphi}_1(t)$ can be explicitly written by 
plugging Eq.~\eqref{sol0} into 
Eq.~\eqref{eq1order}, and taking the Jacobi-Anger expansion, 
\begin{equation}
\begin{split}
\dot{\varphi}_1(t)=&-\frac{1}{2} \sum_{n=-\infty}^\infty\left[I_{2\pi}J_n\left(\frac{I_\text{ac}}{\omega_\text{ac}}\right) \sin((\omega_0-n\omega_\text{ac})t+\phi_0), \right. \\
&\left. + I_{4\pi}J_n\left(\frac{I_\text{ac}}{2\omega_\text{ac}}\right) \sin((\omega_0/2-n\omega_\text{ac})t+\phi_0/2)\right],
\end{split}
\label{shaphigh}
\end{equation}
where $J_n(x)$ is the $n$-th Bessel function.
The time average of Eq.~\eqref{shaphigh} is finite for $\omega_0=n\omega_\text{ac}$, namely
\begin{align}
\langle \dot{\varphi}_1 \rangle=&-\frac{1}{2} \left[I_{2\pi}J_n\left(\frac{I_\text{ac}}{\omega_\text{ac}}\right) \sin(\phi_0) \delta(\omega_0-n\omega_\text{ac}), \right. \nonumber \\
&\left. + I_{4\pi}J_n\left(\frac{I_\text{ac}}{2\omega_\text{ac}}\right) \sin(\phi_0/2)\delta(\omega_0/2-n\omega_\text{ac})\right].
\end{align}
Shapiro steps arise choosing the value of $\phi_0$ that compensates the increment of $I_0$, and thus
$\langle\dot{\varphi}_0\rangle+\langle\dot{\varphi}_1\rangle=n\omega_\text{ac}$ for different values of $I_0$.
Therefore, the step widths will be given by the extreme value of Eq.~\eqref{shaphigh} in respect to $\phi_0$ for the interval $\phi_0=[0,4\pi]$.
Under these approximations, odd and even Shapiro steps
are given by
\begin{equation}
\Delta_{2n-1}=\frac{1}{2}I_{2\pi}\left| J_{2n-1}\left(\frac{I_\text{ac}}{\omega_\text{ac}}\right)\right|,~~~~~~~~~~~~~~~~~~~~~~~
\label{n1}
\end{equation}
\begin{equation}
\begin{split}
\Delta_{2n}=\frac{1}{2} &\text{Max}\left\{ I_{2\pi}J_{2n}\left(\frac{I_\text{ac}}{\omega_\text{ac}}\right)\sin(\phi_0) \right.\\
&~~~~~~~~~~~~~~ \left. + I_{4\pi}J_{n}\left(\frac{I_\text{ac}}{2\omega_\text{ac}} \right) \sin(\phi_0/2) \right\},
\end{split}
\label{n2}
\end{equation}
where $\Delta_n$ is the $n$th-step width given in units of $I_\text{c}$.
In Fig.~\ref{Fig.bessel} we represent $\Delta_n$ for $n=1$ and $n=2$ 
as a function of $I_\text{ac}/\omega_\text{ac}$.
It is important to note that both terms $I_{2\pi}$ and $I_{4\pi}$
enter in the same way in the step widths. 
Therefore, even steps can only dominate for
$I_{4\pi}/I_{2\pi}\gg 1$.
Furthermore, we observe in Fig.~\ref{Fig.bessel} a genuine oscillatory pattern.
Odd step widths show a typical oscillatory pattern, i.e.~they involve only one Bessel function and thus, 
they go to zero for given values of the argument $I_\text{ac}/\omega_\text{ac}$.
In turn, the even step widths are composed by the sum of two different Bessel functions. Thus, 
the step widths show two minima, and none of them reaches zero. 
Therefore, although the even step widths are comparable with 
the odd step widths, the beating pattern of the step widths can 
be used to identify and estimate the intensity of the $4\pi$ component of the supercurrent.

\section{Conclusions}

In this paper we study the dynamics of a Josephson 
junction carrying two superconducting contributions:
a $2\pi$- and a $4\pi$-periodic in phase difference, with intensity $I_{2\pi}$ and $I_{4\pi}$, respectively.
We use the 2S-RSJ model to understand the relation between the dynamics of the junction and 
the width of the Shapiro steps, and in particular we focus on 
the reasons that make the even steps dominate over the odd steps for a fixed ratio $I_{4\pi}/I_{2\pi}\ll 1$. 
This phenomenon \cite{Dominguez2012a} is important because it has been observed in 
different experiments\cite{Rokhinson2012a, Wiedenmann2016a, Deacon2016a}, and could help to determine the 
presence of topological superconductivity. 

We provide a qualitative explanation of this phenomenon in terms of the washboard potential, 
and obtain a phase diagram of the widths of the Shapiro steps as a function of $I_\text{ac}$ and $I_0$.
Remarkably, using some elementary reasonings we find the range 
of AC-bias, i.e.~$I_\text{ac}$, where the non-linear dynamics of the junction 
causes a regime in which the even steps dominate over the odd steps.
Increasing further $I_\text{ac}$ we expect to find a crossover 
to a situation where odd steps are present although even steps dominate. 
Then, at very high values of $I_\text{ac}$, both contributions become 
comparable.

Furthermore, we study analytically the Shapiro step width as a function of $\omega_\text{ac}$ 
in two different limits of $I_\text{ac}$:
the low intensity limit $I_\text{ac}\ll I_\text{c}$, and 
the high intensity limit $I_\text{ac}\gg I_\text{c}$. 
The low intensity limit is precisely the limit where one can find 
only even Shapiro steps even when $I_{4\pi}/I_{2\pi}\ll 1$. 
In this limit,
we find the link between two different experiments: 
the Josephson emission spectrum\cite{Deacon2016a} and the Shapiro 
experiment.\cite{Rokhinson2012a,Wiedenmann2016a,Bocquillon2016a} 
%through the equal periodicities of the time dependent voltage and the inverse 
%of the time dependent voltage, respectively.
In addition, we obtain analytical expressions for the 
step widths in the high intensity limit $I_\text{ac}\gg I_\text{c}$. 
We show that the maximum width of the even and odd Shapiro steps
depends linearly on the ratio of $I_{4\pi}/I_{2\pi}$. 
However, even in this regime one can unravel the existence of the $4\pi$-periodic contribution, 
due to the beating pattern of even Shapiro steps as a function of $I_\text{ac}$.

\begin{figure}[tb]
\begin{center}
\includegraphics[width=3.3in,clip]{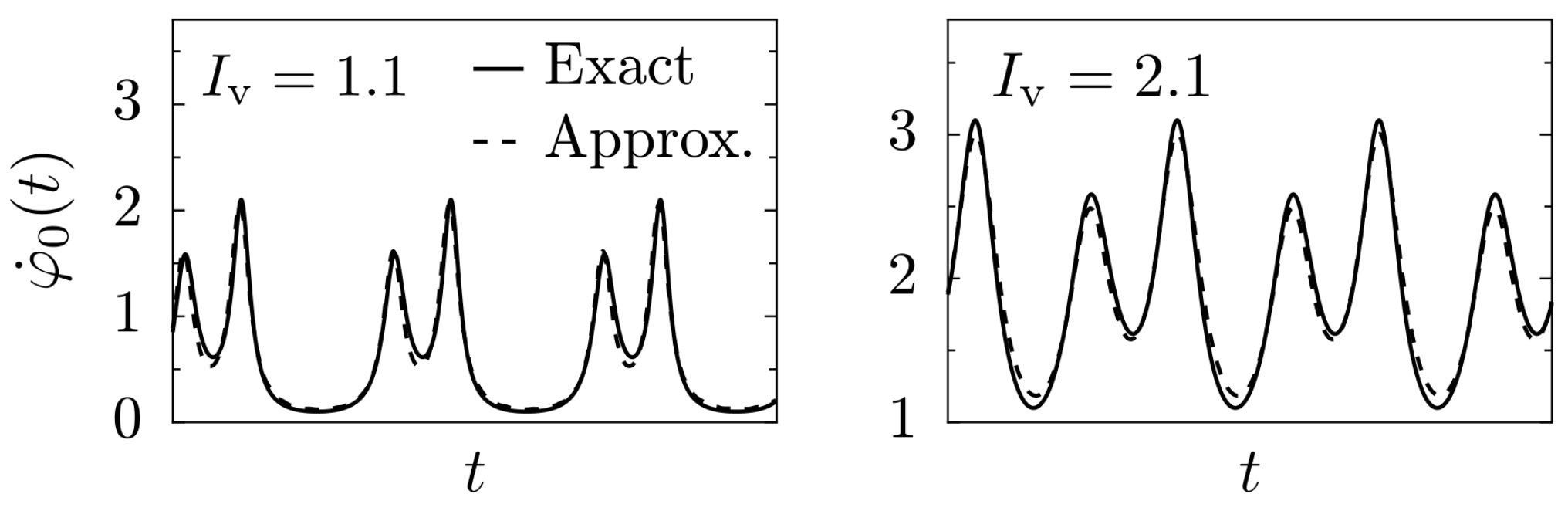}
\end{center}
\caption {\label{Fig.approx}\small 
Comparison between the numerical solution of Eq.~\ref{eomdl} $\dot{\varphi}_0(\tau)$ (solid lines) and the approximate solution given by
Eq.~\eqref{phidot} (dashed lines). We have used $I_{4\pi}/I_{2\pi}=0.5$.
We compare two 
different values of $I_\text{v}=1.1$ (left panel) and $I_\text{v}=2.1$ (right panel).
}
\end{figure}

\acknowledgements
We acknowledge  financial  support  from  the  DFG  via  SFB  1170  "ToCoTronics", 
the Land of Bavaria (Institute for Topological Insulators and the Elitenetzwerk Bayern), the German Research
Foundation DFG (SPP 1666), the European Research Council (advanced grant project 3-TOP), the Helmholtz Association (VITI)
and the Spain's 
MINECO through Grant No. MAT2014-58241-P. 
T.M.K. is financially supported by the European Research Council Advanced grant No.339306 (METIQUM) and by the Ministry of
Education and Science of the Russian Federation under Contract No.14.B25.31.007. T.M.K., E.B. and L.W.M. gratefully thank the Alexander von Humboldt foundation for a Research-prize.
R.S.D. acknowledges support from Grants-in-Aid for Young Scientists B (No. 26790008) and Grants-in-Aid for Scientific Research A (No. 16H02204).
We acknowledge enlightening discussions with 
Y. V. Nazarov, J. Pic\'o, C. Br\"une and H. Buhmann.

\appendix

\section{Adapting the 2$\pi$ solution to the mixed situation}
\label{app.2pi}
The solution of Eq.~\eqref{eomdl} with $I_{4\pi}=0$ and $I_{2\pi}=1$ has been solved previously in 
Ref.~\onlinecite{Aslamazov1969a}, 
\begin{align}
T=\int_0^{2 \pi}\frac{d\varphi}{I_\text{v} - \sin( \varphi ) } =\frac{2 \pi}{\sqrt{I_\text{v}^{2}-1}}
\label{eq.period2pi}
\end{align}
The corresponding frequency $\omega_0=2\pi/T$ is proportional to the voltage. 
Besides, the stationary voltage is equal to the frequency 
$\overline{V}=\omega_0=\sqrt{I_\text{v}^2-1}$. 
In this case the time evolution of $\dot{\varphi}_0(t)$ can be solved exactly and 
is given by 
\begin{align}
\dot{\varphi}_0(t)= \omega_0 \left[1+2\sum_{n=1}^\infty \left( I_\text{v}-\omega_0 \right)^n \cos(n\omega_0 t) \right],
\label{eq.2pi}
\end{align}
for $I_\text{v} > 1$. 
In order to adapt this solution to the more general case, where $I_{4\pi}\neq0$,
we need to take into account the two periods $T_1$ and $T_2$, and also 
to include the different intensities observed in the maxima $F_1,~F_2,~S_1$ and $S_2$ (see Fig.~\ref{Fig.washboard}a).
To this aim, we double the period of the system by substituting $\omega_0$ by $\omega_0/2$, 
with $\omega_0 =4\pi/T_{4\pi}$, and then 
shift the cosine term in two opposite
directions $\pm T_1/2$. In this way 
we tune from a solution that exhibits equally time spaced peaks, where the period $T$ 
is given by Eq.~\eqref{eq.period2pi}, to a function exhibiting 
peaks separated by $T_1$ and $T_2$.
In order to include two periods $T_1$ and $T_2$ mantaining the same height 
one needs to 
renormalize the Fourier coefficients and substitute $( I_\text{v}-\omega_0 )$ 
by its square root of $z=( I_\text{v}-\omega_0 )^{1/2}$,
yielding 

\begin{equation}
\begin{split}
\dot{\varphi}_0(t)\approx & \omega_0 \left[1+\right.\\
& \sum_{n=1}^\infty  z^{n} \left(\cos(n\omega_0 (t+T_1/2)/2)+\right.\\
&\left. \left.\sin(n\omega_0 (t-T_1/2)/2)\right)\right]
\end{split}
\end{equation}
This equation gives rise to peaks exhibiting equal height, 
in order to adjust to the numerical solution, 
we multiply the second term in the sum by $I_{2\pi}$, which 
in the pure $2\pi$ solution was equal to 1, namely
\begin{equation}
\begin{split}
\dot{\varphi}_0(t)\approx & \omega_0 \left[1+\right.\\
& \sum_{n=1}^\infty  z^{n} \left((I_{2\pi}+1)\cos(n\omega_0 T_1/4)\cos(n\omega_0 t/2)+\right.\\
&\left. \left.(I_{2\pi}-1)\sin(n\omega_0 T_1/4)\sin(n\omega_0 t/2)\right)\right]
\end{split}
\label{prevphidot}
\end{equation}
We find that the equation becomes more similar to the numerical results when we substitute
the first coefficient by 2, that is, $(I_{2\pi}+1)\rightarrow 2$, yielding the result 
given in Eq.~\eqref{phidot}. 
In Fig.~\ref{Fig.approx} we show how accurate the approximate solution is, 
by comparing it against the numerical result.

\end{document}